\documentclass[aoas,MSNbibl,nameyear,dvips]{arximspdf}
\usepackage{graphicx}
\doi{10.1214/11-AOAS489}
\volume{5}
\issue{4}
\pubyear{2011}
\firstpage{2470}
\lastpage{2492}



\begin{document}
\begin{frontmatter}

\title{Efficient emulators of computer experiments using compactly
supported correlation functions, with an application to cosmology}
\runtitle{Efficient emulators of computer experiments}

\begin{aug}
\author[A]{\fnms{Cari G.} \snm{Kaufman}\corref{}\thanksref{t1}\ead
[label=e1]{cgk@stat.berkeley.edu}},
\author[B]{\fnms{Derek}
\snm{Bingham}\thanksref{t2}\ead[label=e2]{dbingham@stat.sfu.ca}},
\author[C]{\fnms{Salman}
\snm{Habib}\thanksref{t3}\ead[label=e3]{habib@lanl.gov}},\\
\author[D]{\fnms{Katrin} \snm{Heitmann}\thanksref{t3}\ead
[label=e4]{heitmann@lanl.gov}} and
\author[E]{\fnms{Joshua A.} \snm{Frieman}\ead[label=e5]{frieman@fnal.gov}}
\runauthor{C. G. Kaufman et al.}
\affiliation{University of California, Berkeley, Simon Fraser
University, Argonne National Laboratory, Los Alamos National
Laboratory, and Fermi National Accelerator Laboratory and University of
Chicago}
\address[A]{C. G. Kaufman\\
Department of Statistics\\
University of California, Berkeley\\
Berkeley, California 94720\\
USA\\
\printead{e1}}
\address[B]{D. Bingham\\
Department of Statistics\\
\quad and Actuarial Science\\
Simon Fraser University\\
V5A 1S6\\
Burnaby, BC\\
Canada\\
\printead{e2}\hspace*{32.29pt}}
\address[C]{S. Habib\\
High Energy Physics Division\\
Mathematics and Computer Science Division\\
Argonne National Laboratory\\
Argonne, Illinois 60439\\
USA\\
\printead{e3}}
\address[D]{K. Heitmann\\
ISR-1, ISR Division\\
Los Alamos National Laboratory\\
Los Alamos, New Mexico 87544\\
USA\\
\printead{e4}}
\address[E]{J. Frieman\\
Center for Particle Astrophysics\\
Fermi National Accelerator Laboratory\\
Batavia, Illinois 60510\\
USA\\
\printead{e5}}
\end{aug}

\thankstext{t1}{Supported in part by   NSF Grant DMS-06-35449 to
the Statistical and Applied Mathematical Sciences Institute.}

\thankstext{t2}{Supported in part by a grant from the
Natural Sciences and Engineering Research Council of Canada.}

\thankstext{t3}{Supported in part by the Department of
Energy under contract W-7405-ENG-36.}

\received{\smonth{6} \syear{2010}}
\revised{\smonth{4} \syear{2011}}

%
\begin{abstract}
Statistical emulators of computer simulators have
proven to be useful in a variety of applications. The widely adopted
model for emulator building, using a Gaussian process model with
strictly positive correlation function, is computationally
intractable when the number of simulator evaluations is large. We
propose a new model that uses a~combination of low-order regression
terms and compactly supported correlation functions to recreate the
desired predictive behavior of the emulator at a fraction of the
computational cost. Following the usual approach of taking the
correlation to be a product of correlations in each input dimension,
we show how to impose restrictions on the ranges of the correlations,
giving sparsity, while also allowing the ranges to trade off against
one another, thereby giving good predictive performance. We
illustrate the method using data from a computer simulator of
photometric redshift with 20,000 simulator evaluations and 80,000
predictions.
\end{abstract}

%

\end{frontmatter}

Simulation of complex systems has become commonplace in scientific
studies. Frequently, simulators (or computer models) are
computationally demanding, and relatively few evaluations can be
performed. In other cases, the computer models are fast to evaluate
but are not readily available to all scientists. This arises, for
example, when the simulator runs only on a~supercomputer or must be
run by specialists. In either case, a statistical emulator of the
computer model can act as a surrogate, providing predictions of the
computer model output at unsampled inputs values, with corresponding
measures of uncertainty [see, e.g., \citet{Sacks1989},
\citet{Santner2003}]. The
emulator can serve as a component in probabilistic model calibration
[\citet{Kennedy2001}], and it can help provide insight into the
functional form of the simulator response and the importance of
various inputs [\citet{Oakley2004}].

Building an emulator can be viewed as a type of nonparametric
regression problem, but with a key difference. Computer experiments
differ from their real-world counterparts in that they are typically
deterministic. That is, running the code twice with the same set of
input values will produce the same result. To deal with this
difference from the usual noisy settings, \citet{Sacks1989} proposed
modeling the response from a computer experiment as a realization from
a Gaussian process~(GP). From a Bayesian viewpoint, one can think of
the GP as a prior distribution over the class of functions produced by
the simulator.

\begin{figure}
\begin{tabular}{@{}c@{\qquad}c@{}}

\includegraphics{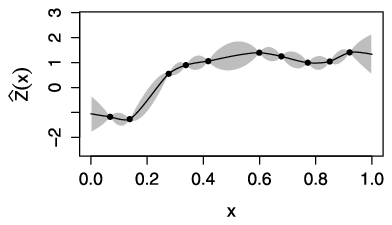}
 & \includegraphics{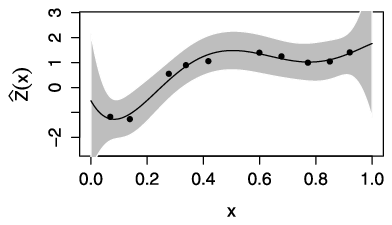}\\
(a) & (b)\\[4pt]

\includegraphics{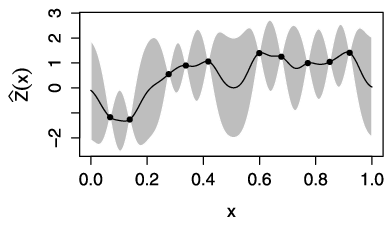}
 & \includegraphics{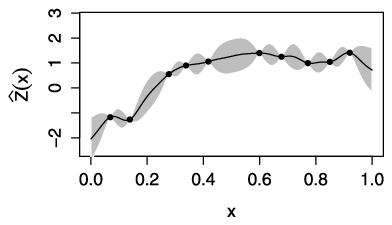}\\
(c) & (d)
\end{tabular}
\caption{Illustrative example in which the data were drawn from a mean
zero Gaussian process with covariance function $K(x, x') =
\exp\{-5|x-x'|^{1.5}\}$ and predictions were made using one of four
methods. Each plot shows the observations (solid dots), predictions
(solid line) and pointwise 95\% confidence intervals for the
predictions (gray bands). The models were \textup{(a)} zero mean GP
with the correct covariance structure, \textup{(b)} OLS using Legendre
polynomials up to degree 6, \textup{(c)} zero mean GP with the Bohman
correlation function (\protect\ref{bohman}), with $\tau=0.1$, and
\textup{(d)} GP with Legendre polynomials up to degree 2 in the mean
and Bohman correlation function with $\tau=0.1$.} \label{figtoyexample}
\end{figure}

The GP model is particularly attractive for emulation because of its
flexibility to fit a large class of response surfaces. It is also
desirable that the statistical model, and corresponding prediction
intervals, reflect some type of smoothness assumption regarding the
response surface, leading to zero (or very small) predictive
uncertainty at the design points, small predictive uncertainty close
to the design points, and larger uncertainty further away from the
design points. For example, note the behavior of the confidence
intervals in the illustration shown in panel (a) of
Figure \ref{figtoyexample}.

%

The main drawback of the usual GP model in our setting is that it can
be computationally infeasible for large designs. The likelihood for
the observations is multivariate normal, and evaluation of this
density for $n$ design points requires manipulations of the $n \times
n$ covariance matrix
that grow as~$O(n^3)$. This limitation is the main motivation for this
article, which was prompted by our work on just such an application in
cosmology (see Section~\ref{example}). Here, the basic idea is to
construct a statistical model based on a~set of simulated
data consisting of multi-color photometry for training
set galaxies, as well as the corresponding true redshifts. Given
photometric information for a test galaxy, the system should produce
an estimated value for the true redshift. The very large experimental
design used to explore the input space, that is, the large number of
galaxies used to build the training set, presents a computational
challenge for the GP.

It is worth noting that the GP model is not the only approach for
emulating computer simulators.
Using a GP with constant mean term can be viewed as a way of forcing
the covariance structure of the GP to model all the variability in
the computer output. At the other extreme, one can take a regression
based approach
and treat the errors as white noise, as in \citet{An2001}. This
approach has the benefit of being computationally efficient, as the
correlation matrix is now the identity matrix. However, it does not
have the attractive property
of a smooth GP, namely, that the predictive distribution interpolates
the observed data and that the uncertainty reflects the above
properties. Instead, the white noise is introducing random error to
the problem that is not actually believed to exist; it is there simply
to reflect lack of fit. The implications of this for predictive
uncertainty are illustrated in panels (a) and (b) of Figure
\ref{figtoyexample}. Panel (a) shows a set of data fit using the
standard GP model, and panel (b) shows the same data fit using
ordinary least squares regression with the set of Legendre polynomials
up to degree six. The behavior in panel (a) is what we desire in an
emulator, but the model is computationally intractable for large data
sets. The model in panel (b) is very efficient from a computational
standpoint, but the predictions do not reflect the determinism of the
computer simulator. The approach proposed in this article can be
viewed as finding an intermediate model to those in panels (a) and
(b), such that the model is both fast to fit and has the appropriate
behavior for prediction.

In this article, we propose new methodology for emulating computer
simulators when $n$ is large (i.e., when it is infeasible to fit the
usual GP model). The approach makes the key innovation of
replacing the covariance function with one that has compact support.
This has the effect of introducing zeroes into the covariance matrix,
so that it can be efficiently manipulated using sparse matrix
algorithms. In addition, the proposed approach easily handles the
anisotropy that is common in computer experiments by allowing the
correlation range in each dimension to vary and also imposes a
constraint on these ranges to enforce a minimum level of sparsity in
the covariance matrix. We further propose a model for the mean of the
GP, rather
than taking it to be a scalar. The motivation for this is that the
introduction of regression functions
tends to decrease the estimated correlation length in the GP,
thereby offsetting some of the loss of predictive efficiency
introduced by using a compactly supported covariance. Last, we
propose prior distributions to incorporate experimenter belief and also
to make the application of regression terms and the compactly supported
covariance function efficiently work together.

In the next section we introduce the GP that is commonly used for
building emulators and illustrate the challenges for large
data sets. In Section~\ref{implementation} we present new methodology for building
computationally efficient emulators, and we give some details of the
implementation and computational advantages in Section \ref{simulation}.
In Section
\ref{example} we investigate the performance of the method in a simulation study.
The method is
then used to construct an emulator of photometric redshifts
of cosmological objects in Section \ref{sec6}, and we conclude with some final
remarks in the \hyperref[app]{Appendix}.

\section{Gaussian process models for computer experiments}\label{gpmod}

Consider a simulator that takes inputs $\mathbf x \in\Re^d$ and produces
univariate output $Y(\mathbf x)$. The GP model generally used in this
setting treats the response as a realization of a random function:
%
%
\begin{equation}\label{surface}
Y({\mathbf x}) = \sum_{i=1}^p f_i({\mathbf x})\beta_i + Z({\mathbf x}),
\end{equation}
where $f_1, \ldots, f_p$ are fixed regression functions, ${\bolds
\beta} = (\beta_1, \ldots, \beta_p)'$ is a vector of unknown
regression coefficients, and $Z$ is a mean zero GP. The covariance
function of $Z$ is denoted by
%
%
\begin{equation}\label{covfunction}
\operatorname{Cov}(Z({\mathbf x}), Z({\mathbf x}')) = K({\mathbf x}, {\mathbf x}'; \sigma^2,
\bolds\theta) = \sigma^2 R({\mathbf x}, {\mathbf x}'; \bolds\theta),
\end{equation}
where $\sigma^2$ is the marginal variance of $Z$
and $\bolds\theta$ is a vector of parameters controlling the
correlation.

We defer until the end of this section a discussion of the choice of
$f_1, \ldots, f_p$ and $R$ and first lay out some general notation.
Let $\mathbf{Y} = (Y(\mathbf{x}_1), \ldots, Y(\mathbf{x}_n))'$ be the vector of
observed responses. Then, ignoring\vspace*{1pt} a constant, the log-likelihood
under this model is $l({\bolds\beta}, \bolds
\theta,\sigma^2) = - {\frac{1}{2}\log}|\mathbf{R}(\bolds\theta)| -
\frac{1}{2\sigma^2} (\mathbf{Y} -\break \mathbf{F}{\bolds\beta})'
\mathbf{R}(\bolds\theta)^{-1} (\mathbf{Y} - \mathbf{F}{\bolds\beta})$,
where $\mathbf{F}$ is the $n \times p$ matrix of regression functions and
$\mathbf{R}(\bolds\theta)$ is the $n \times n$ matrix of
correlations with $[\mathbf{R}(\bolds\theta)]_{ij} = R(\mathbf{x}_i,
\mathbf{x}_j; \bolds\theta)$.
In addition, for any set of model parameters, the conditional
distribution of $Y(\mathbf{x}_0)$ at a new input value, $\mathbf{x}_0$, given the observations $\mathbf{Y}$, is normal with mean and
variance
%
%
\begin{eqnarray}
\label{predictmean}
E[Y(\mathbf{x}_0) | \mathbf{Y}, \bolds\beta, \sigma^2, \bolds
\theta] &=&
f(\mathbf{x}_0) {\bolds\beta} + \mathbf{r}_0(\bolds
\theta)' \mathbf{R}(\bolds\theta)^{-1}(\mathbf{Y} - \mathbf{F}{
\bolds\beta}),\\
\label{predictvar}
\operatorname{Var}[Y(\mathbf{x}_0) | \mathbf{Y}, \bolds\beta, \sigma^2, \bolds
\theta] &=&
\sigma^2[1 - \mathbf{r}_0(\bolds\theta)' \mathbf{R}(\bolds
\theta)^{-1} \mathbf{r}_0(\bolds\theta)],
\end{eqnarray}
where $\mathbf{r}_0(\bolds\theta)$ is the $n$-vector of
correlations between the observed responses and $Y(\mathbf{x}_0)$.

In practice, $\bolds\beta$, $\sigma^2$ and $\bolds\theta$
are unknown and must be estimated.
This can be achieved using likelihood-based methods such as maximum
likelihood (ML) or restricted maximum likelihood (REML) [see, e.g.,
\citet{Irvine2007} for a comparison of ML and REML estimation].
Alternatively, one may specify a Bayesian model in which the joint
posterior distribution for both parameters and predicted values of the
function can be approximated via
Markov Chain Monte Carlo (MCMC).
We adopt the latter approach, although most of the proposed
methodology is also applicable in a frequentist setting.

Two choices must be made to complete the specification in
(\ref{surface}) and (\ref{covfunction}), the regression functions
$f_1,\ldots,f_p$ and the correlation function $R$. The mean structure
in (\ref{surface}) is almost always taken to be flat over the domain
of the function, with \mbox{$f(x) \equiv1$}.
By far the most common specification for the correlation function
is a product of one-dimensional power exponential correlation
functions. Specifically, writing $\bolds\theta_k = \{\phi_k,
\alpha_k\}$,
%
%
\begin{eqnarray}
\label{prodcov}
R({\mathbf x}, {\mathbf x}'; \bolds\theta) &=& \prod_{k=1}^d R_k (|
x_{k} - x_{k}' | ; {\bolds\theta}_k) \qquad\mbox{(Product
form)}\\
\label{powerexp}
&=& \prod_{k=1}^d \exp\{ -\phi_k | x_{k} - x_{k}' | ^{\alpha_k} \}
\qquad\mbox{(Power exponential)}
\end{eqnarray}
for $\phi_k > 0$ and $1 \leq\alpha_k \leq2$. Since the ${\bolds
\phi}_k$'s are not constrained to be equal, the model can handle
different signals in each input dimension of the simulator (i.e.,
anisotropy). This choice of constant mean term and power exponential
correlation is so common in practice that we will refer to it as the
``standard model.'' In the next section we shall diverge from the
standard model and propose a more computationally efficient model with
different mean and covariance structures.

No matter what choices are made for the regression terms and
correlation function, inference and prediction for this model requires
evaluation of the log-likelihood,
typically many times. These calculations require finding the log
determinant of $\mathbf{R}(\bolds\theta)$ and solving $\mathbf{R}
(\bolds\theta)^{-1} (\mathbf{Y} - \mathbf{F}{\bolds\beta})$.
When the correlation functions are strictly positive,\vspace*{1pt} as in the
standard model, both of these grow in complexity by order $n^3$, and
therein lies the problem. These operations are intractable for
moderate sample sizes and simply impossible for large sample sizes. It
is this problem that motivates the current work.

\section{Building computationally efficient emulators}\label{sec2}

Our approach is conceptually straightforward, consisting of three main
innovations:
\begin{longlist}[(2)]
\item[(1)] using compactly supported correlation functions to model
small-scale variability,
\item[(2)] using regression functions in the mean of the GP to model
large-scale variability, and
\item[(3)] specifying prior distributions for model parameters (or parameter
constraints, in the frequentist case) that are
flexible while still enforcing computational efficiency.
\end{longlist}
These three innovations work together to produce a flexible, fast and
powerful method for computer model emulation.

\subsection{Compactly supported correlation functions}\label{compact}

We begin by proposing that the correlation functions in the product
covariance (\ref{prodcov}) are chosen to have compact support. That
is, for some $\tau_k \geq0$, \mbox{$R_k(| x_{k} - x_{k}'| ;\tau_k)=0$}
when  $| x_{k} - x_{k}'| \geq\tau_k$. This has the effect of
introducing zeros into the covariance matrix, so computationally
efficient sparse matrix techniques [\citet{Pissanetzky1984},
\citet{Barry1997}]
may be used.
Compactly supported correlation functions have recently received
increased attention in the literature, used either by themselves
[\citet{Gneiting2002}, \citet{Stein2008}] or multiplying
another, strictly positive
correlation function, which is known as tapering
[\citet{Furrer2006}, \citet{Kaufman2008}]. These applications
all assume that the
compactly supported correlation function is isotropic, requiring a
single range parameter. In contrast,
anisotropic covariance functions are the norm for computer experiments
because the inputs to the computer model are frequently on different
scales and/or impact the response in vastly different ways. Therefore,
we use correlation functions with compact support in product form.

We focus on two families of models that can be used to approximate the
power exponential function (\ref{powerexp}). These functions are zero
for $t \geq\tau$, and for $t < \tau$,
%
%
\begin{eqnarray}
\label{bohman}
\mbox{Bohman:}\hspace*{22pt}\quad R(t; \tau) &=& (1-t/\tau)\cos(\pi t/\tau) + \sin(\pi
t/\tau)/\pi,\\
\label{truncpow}
\qquad\mbox{Truncated power:}\quad R(t;\tau,\alpha,\nu) &=&
[1-(t/\tau)^\alpha]^\nu,\nonumber\\[-8pt]\\[-8pt]
&&\eqntext{0 < \alpha< 2, \nu\geq\nu_d(\alpha).}
\end{eqnarray}
The term $\nu_d(\alpha)$ in (\ref{truncpow}) represents a restriction
so that the function is a valid correlation, with $\lim_{\alpha
\rightarrow2} \nu_d(\alpha) = \infty$ [\citet{Golubov1981}]. Although
it is difficult to calculate $\nu_d(\alpha)$ exactly,
\citet{Gneiting2001} gives upper bounds for a variety of values of
$\alpha$ between 1.5 and 1.955. For example, upper bounds for~$\nu
_1(\alpha)$ when $\alpha= 3/2$ and $5/3$ are 2 and 3, respectively.

These two functions allow for some flexibility in the smoothness of
the realizations of the GP, in loose analogy to the parameter $\alpha$
in the power exponential model~(\ref{powerexp}). The truncated power
function (\ref{truncpow}) is not differentiable even once at the
origin, with $\alpha< 2$, corresponding to a process which is not
even mean square continuous, as for (\ref{powerexp}) with $\alpha<
2$. In contrast, the Bohman function (\ref{bohman}) is twice
differentiable at the origin, corresponding to a~process which is mean
square differentiable. Of course, when $\alpha= 2$, the power
exponential function (\ref{powerexp}) is infinitely differentiable at
the origin.
However, we feel this level of smoothness is often unrealistic, and in
our applied work we have not found any reason to prefer a higher level
of smoothness than is given by~(\ref{bohman}).

Note that the ranges play two different roles in our approach. First,
they control the degree of correlation in each dimension. In this way
they are similar to the range parameters $\phi_k$ in the power
exponential function (\ref{powerexp}). However, unlike the~$\phi_k$,
the $\tau_k$ also control the degree of sparsity of the correlation
matrix. To produce computational savings, some restrictions must be
applied to the $\tau_k$. We chose to do this through the prior
distribution, which we discuss in Section \ref{priors}.

\subsection{Regression functions}

We propose using regression
functions to mo\-del the mean structure of a computer model, rather than
the constant mean used in the standard model. The intuition behind our
approach is that if the large-scale structure of the simulator output
is well captured by a linear combination of the basis functions $f_i$,
we would naturally expect the residual process $Z({\mathbf x})$ to have
shorter-range correlations.
This type of trade-off between large scale and small scale variability
has been noted in the spatial statistics literature [\citet{Cressie1993},
\citet{Stein2008}], and it was noted in the results of a simulation
study of computer experiments by \citet{Welch1992}. However, to our
knowledge, it has not previously been exploited for computational
purposes in constructing emulators, an application in which we will see
the often high dimensionality of the input ${\mathbf x}$ allows this
trade-off to be leveraged particularly well.

Using a richer mean structure produces more efficient predictions than
the use of the compactly supported covariance alone.
For example, panel~(c) of Figure \ref{figtoyexample} illustrates the
predictions and pointwise confidence bands under the zero mean GP
model with Bohman correlation function (\ref{bohman}) with $\tau=0.1$.
These are obviously less efficient than the results under the true
correlation function
in panel (a).
However, this limitation can be addressed by incorporating regression terms.
Panel (d) in Figure \ref{figtoyexample} illustrates the results of
combining a small set of basis functions with a compactly supported
correlation function. The behavior of the predictions is similar to
that in panel~(a), but the model in panel (d) can be applied to large
data sets, whereas the model in (a) cannot.

There are a variety of basis functions among which one can choose, and
detailing them is not the focus of this article. We have found a good
default choice to be taking $f_i$ to be tensor products of Legendre
polynomials over $[0,1]$, the input variable in each dimension having
been rescaled to this domain [see, e.g., \citet{An2001}].

\subsection{Prior distributions} \label{priors}

The full specification of our Bayesian model requires that we choose
prior distributions for the parameters $\bolds\tau$, $\sigma^2$
and $\bolds\beta$. We advocate the inclusion of prior
information where available, but describe here a default choice of
prior distributions that brings additional computational efficiency.
These advantages may be had in a frequentist setting by replacing the
prior distributions with certain restrictions on the parameter space,
which should be obvious as we proceed.

One of the novel features of the proposed approach is use of an
anisotropic compactly supported covariance and the estimation of the
range parameters. Recall that the $\tau_k$'s play the role of
governing the correlation in each dimension and also controlling the
computational complexity of the model. Unfortunately, the mapping from
a collection $\tau_1, \ldots, \tau_d$ to the sparsity of a~given
matrix will depend on the particular configuration of sampling points,
and it is also difficult to translate a degree of sparsity (measured,
say, by the percentage of off-diagonal zeroes) to computation time for
a particular algorithm. However, we have found that with a bit of
trial and error, controlling the sparsity of the correlation matrix
$\mathbf{R}$ can be an adequate proxy for controlling computation time.
The question then is, how do we control sparsity through the prior for
${\bolds\tau}$?

Throughout, we will assume that all input variables have been scaled
to $[0,1]$, so that taking $\tau_k > 1$ for all $k$ introduces no
sparsity. A simple restriction is
%
%
\begin{equation}\label{cube}
\tau_j \leq B \qquad\mbox{for all }j, B>0.
\end{equation}
However, using this restriction in creating a prior distribution
ignores an important advantage of using compactly supported
correlation functions within the product correlation given in
(\ref{prodcov}): that $R(x, x') \geq0$ only if $|x_j-x'_j| < \tau_j$
for all $j=1,\ldots,d$. That is, a pair of observations having zero
correlation in \textit{any} dimension will be independent and contribute
a zero to the overall correlation matrix. Therefore, the $\tau_j$ may
trade off against one another to produce the same degree of sparsity.
Therefore, we restrict $\tau$ further by taking $\tau$ to be uniformly
distributed over the space
%
%
\begin{equation}\label{triangle}
T_C = \Biggl\{\tau\in\Re^d\dvtx\tau_j \geq0\mbox{ }\forall d,
\sum_{j=1}^d \tau_j \leq C\Biggr\},\qquad C > 0.
\end{equation}
This can be thought of as defining the prior distribution over a
$d$-dimensio\-nal triangle rather than a cube. Because of the trade-off
between the $\tau_j$ discussed above, $C$ can generally be greater
than $B$ is in (\ref{cube}) and impose the same degree of sparsity.
That is, some of the $\tau_j$ are allowed to be large, which they may
indeed need to be to reflect a high degree of correlation in
particular input dimensions. Some trial and error may be required to
find a $C$ for which calculations can actually be carried out; we
return to this in the next section. This restriction may not work
well for the case that certain input variables have no impact on the
output variable
over the range being simulated, which would correspond to $\tau_j
\rightarrow\infty$. We
make the assumption that such variables have been previously identified
and fixed during
a prior screening analysis such as described in \citet{Welch1992}
or \citet{Linkletter2006}.\looseness=-1

Finally, we specify prior distributions for $\sigma^2$ and $\bolds
\beta$. We follow Berger, De Oliveira and Sans{\'o}
(\citeyear{Berger2001}) and \citet{Paulo2005}, who proposed the
form $p(\sigma^2,\beta,\tau) \propto\pi(\tau)/\sigma^2$ for
Gaussian processes. As our choice of $\pi$ is a proper density, it can
be shown that this choice still produces a proper
posterior\vspace*{1pt} [\citet{Berger2001}]. One advantage of this
choice is that $\beta$ and $\sigma^2$ may be easily integrated out of
the model, so that posterior sampling may be done only over the vector
$\tau$.

We end this section with a note relating our proposed model to existing
works in the field of spatial statistics, in which estimation and
prediction under large sample sizes has seen much recent development.
These approaches may be characterized broadly as either approximating
the likelihood for the original model, or changing the model itself to
one that is computationally more convenient. Examples of the former
include \citet{Stein2004} and \citet{Kaufman2008}, while a
common modification to the model itself is to represent the random
field in terms of a lower-dimensional random variable; models falling
under this framework are reviewed by \citet{Wikle2010}.
The approach we take in this paper is to change the model rather than
approximate it. We did consider using compactly supported correlation
functions as an approximation tool rather than using them directly,
following \citet{Kaufman2008}. Ultimately, however, since there is
no ``true'' random process generating the output of the computer
simulator, we decided to take the conceptually simpler approach of
modifying the model itself. That is, in this nonparametric regression
context, the GP is simply a tool for expressing prior beliefs about the
function, and this may be done effectively using compactly supported
correlation functions when regression terms are also included in the model.

\section{Implementation and computational considerations}\label{implementation}

Implementation of these methods for a given data set proceeds in two
steps. The first step is to sample from the posterior distribution
$p(\bolds\tau| \mathbf{Y})$ using a Metropolis sampler. As
noted above, our choice of prior distribution allows us to work with
this marginal distribution, integrating out $\sigma^2$ and $\bolds
\beta$.
This leads to additional computational savings, as we can sample the
vector $\bolds\tau$ using
an efficient, adaptive Metropolis sampler, the details of which are
described in the \hyperref[app]{Appendix}. The second step is to use
these samples to
generate predictions at new input values. Rather than incorporating
this into the sampler, we recommend another computational trick here,
which is to calculate conditional means and variances at a subset of
the iterations (i.e., conditional on~$\bolds\tau^{(i)}$) and
use these to reconstruct the predictive mean and variance using laws
of iterated expectation. The details of this calculation are also
described in the \hyperref[app]{Appendix}. In the remainder of this
section, we
outline the computational savings to be gained from our approach,
compared to standard, nonsparse techniques.

The demanding aspects of the calculations all involve $\mathbf{R}
(\bolds\tau)$, the correlation matrix for a particular value of
$\bolds\tau$. To efficiently calculate the quantities involved, after
we propose a new $\bolds\tau$ in the Metropolis step, we first compute
a sparse representation of $\mathbf{R} (\bolds\tau)$, then compute the
quantities which will be used in calculating the integrated likelihood.
Here, we use the \texttt{spam} package in \texttt{R}
[\citet{Furrer2010}], which uses the efficient ``old Yale sparse
format'' for representing sparse matrices, where only the nonzero
elements, column indices and row pointers are stored. From a
computational viewpoint, operations involving the zero elements need
not be performed. The time-consuming steps in evaluating the likelihood
are then as follows:
\begin{longlist}[(2)]
\item[(1)] Identifying pairs of input values ${\mathbf x}_i$ and ${\mathbf x}_j$ such
that $|x_{ik} - x_{jk}| < \tau_k$, $\forall k = 1, \ldots, d$. (All
other pairs will contribute a zero to the matrix.)
\item[(2)] For only
these pairs, computing $\prod_{k=1}^d R_k(|x_{ik}-x_{jk}|; \tau_k)$
and using only these to create the sparse representation of $\mathbf{R}
(\bolds\tau)$.
\item[(3)] Computing the
Cholesky decomposition of the sparse matrix object, that is,
$\mathbf{Q}(\bolds\tau)$ such that $\mathbf{R}(\bolds\tau) =
\mathbf{Q}(\bolds\tau)' \mathbf{Q}(\bolds\tau)$.
\item[(4)] Backsolving to compute $(\mathbf{Q}(\bolds\tau)')^{-1}
\mathbf Y$ and $(\mathbf{Q}(\bolds\tau
)')^{-1}\mathbf F$.
\end{longlist}

\begin{figure}

\includegraphics{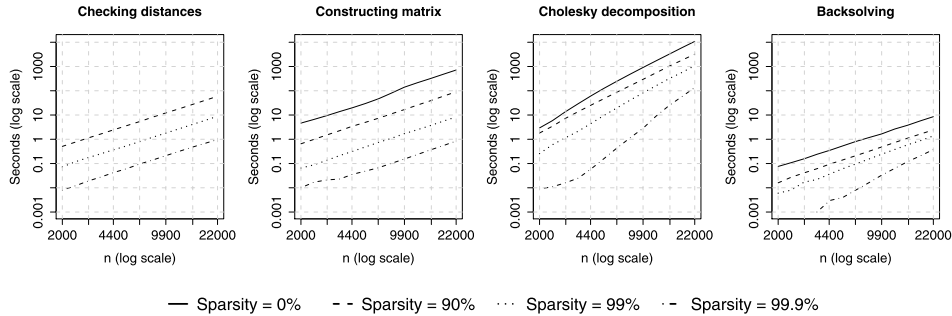}

\caption{Number of seconds required to carry out each step in
evaluating the likelihood, for varying degrees of sparsity and
sample sizes. The case ``$\mbox{Sparsity} = 0$\%,'' shown as a solid line,
corresponds to using a strictly positive correlation function,
while for the other cases ``Sparsity'' denotes the percentage of
off-diagonal elements in the matrix that are equal to zero. The
``Checking distances'' step only applies to the algorithm used
for compactly supported correlation functions, which is why no
solid line appears in this plot. All measurements are shown on
the log (base 10) scale, and each gridline along the y-axis
corresponds to one order of magnitude.}
\label{figsimtimingplot}
\end{figure}

Figure \ref{figsimtimingplot} shows the average number of seconds
required to carry out these steps for a reference data set consisting
of locations uniformly sampled over the input space $[0,1]^4$, which
%
%
is the same dimension as our example in Section~\ref{example}. The
calculations were carried out for various sample sizes and varying
degrees of sparsity, which is imposed by the choice of the cutoff $C$.
Each calculation was repeated 10 times, and the number of seconds
required for each were averaged to produce the plot. Computations were
performed on a~Dell quad socket machine with four dual-core AMD
Opteron 2.4 GHz CPUs and 8 GB of RAM.
All axes are on the log scale. In particular, each gridline along the
y-axis corresponds to one order of magnitude. Overall, we can see that
using a sparse correlation function reduces most of the calculations
by one to three orders of magnitude.

\section{Simulation study}\label{simulation}

The method we present here is intended for use in situations in which
the ``standard'' method is computationally infeasible. However, it is
instructive to see how our method compares when $n$ is small enough
that the standard method can actually be implemented.
In this section we compare the performance of our proposed model to
that of the standard model, when the data are actually simulated under
the standard model. Of course, in practice, neither model is
``correct,'' since the object of interest is a deterministic function,
not a realization of a stochastic process, but this framework allows us
to compare the efficiency of predictions made under varying degrees of
smoothness and for different correlation lengths.

In carrying out the simulation study, we vary both the distribution of
the data and the choices made in fitting it. Regarding the data, we
consider processes in two or four dimensions, with correlation
function (\ref{powerexp}) with $\alpha_k$ set equal (in all
dimensions) to 1.5 or 1.99. (We do not consider the infinitely smooth
case of $\alpha= 2$ here, to avoid dealing on a case-by-case basis
with the numerical instabilities that sometimes arise.) We specify
$\phi_k$ in (\ref{powerexp}) to be such that the effective range,
defined as the distance beyond with correlations are less than 0.05,
is either 0.5 or 2 in all dimensions. This gives six possible
combinations.

For each combination and each of 100 replications of the simulation, we
consider sample sizes of $n=100,150,250,400,650$ and $1\mbox{,}100$. The design
(choice of input settings) for each $n$ and replication is generated
using a random Latin hypercube sample (LHS)
[\citet{McKay1979}] on $[0,1]^d$. In addition, we generate a set
of $n_{\mathrm{pred}} = 512$ input settings at which to predict the function,
using the orthogonal array-based Latin hypercube sampling (OA-LHS)
algorithm by \citet{Tang1993}, with frequency $\lambda=2$. We
sampled values of $Y(x)$ over the LHS and OA-LHS jointly, corresponding
to a realization from the GP.
The rationale for this sampling strategy stems from our different
priorities when choosing the design points and choosing the prediction
points. Repeated sampling under simple LHS is our attempt to mimic the
broad class of designs experimenters mights use in practice. To best
evaluate the predictive accuracy of our methods over that class,
however, we use OA-LHS to choose the evaluation points, due to its
superiority over simple LHS in integral approximation.

In fitting each data set and making the predictions, we either use the
standard model with the $\alpha_k$ set to the value used in simulating
the data, or we use the method outlined in the previous section. For
the latter, we have some choices to make. The first is the basis
functions $f_i$ to include in (\ref{surface}). We use fifth order
Legendre polynomials in each dimension. Specifically, we include all
main effects up to order five, as well as all interactions involving
two dimensions, in which the sum of the maximum power of the exponent
in each dimension is constrained to be less than or equal to $5$. For
example, in two dimensions, this implies using terms involving $x_1,
\ldots, x_1^5, x_2, \ldots, x_2^5, x_1x_2, x_1x_2^2, \ldots, x_1^4x_2$.
We have investigated the impact of the order of the
polynomial for the regression functions and have found that fourth or
fifth order polynomials are sufficient for most applications. We do not
try here
to adapt the order of the polynomial to each particular data set. In
practice, we recommend routinely using a subset of the data and
performing ordinary least squares regression for increasing degrees of
the polynomial. The smallest degree of the polynomial where the fit is
satisfactory is chosen to be the order used in the approach (more on
this in Section \ref{sec51}).

The final choice is the value of the cutoff $C$ in (\ref{triangle}).
Here we chose $C$ such that the maximum proportion of off-diagonal
elements was either 0.02 or 0.05, both of which reflect a high degree
of sparsity, as we would often be required to specify in practice. The
truncated power correlation function (\ref{truncpow}) is used; the
Bohman function (\ref{bohman}) gives very similar results.

The predictions are compared using two criteria. The first,
sometimes called the Nash--Sutcliffe efficiency,
is equal to
%
%
\begin{equation}\label{lof}
\mathrm{NSE} = 1-\frac{\sum_{x \in\mathcal{X}_{\mathrm{pred}}} (\hat{Y}(x) -
Y(x))^2}{\sum_{x \in\mathcal{X}_{\mathrm{pred}}} (Y(x)-\bar{Y})^2}.
\end{equation}
Here, $\hat{Y}(x)$ represents the mean of the posterior predictive
distribution for~$Y(x)$ given the vector of observations $\mathbf{Y}$,
while $\bar{Y}$ represents the mean of~$\mathbf{Y}$. Predictions are made
at the set of 500 hold out points, $\mathcal{X}_{\mathrm{pred}}$. The second
term of (\ref{lof}) is the ratio of an estimate of the predictive mean
square error to the unstandardized\vadjust{\goodbreak} variance of $Y(x)$. Thus, the NSE
has an interpretation similar to $R^2$ in linear regression, insofar
as it represents an estimate of the proportion of the variability in
$\mathbf{Y}$ that is explained by the model, although measured on an
out-of-sample test set. We estimate the average value
of (\ref{lof}) across repeated samples by calculating it for each
iteration of the simulation study and then averaging over iterations.
We do not expect NSE to be better (larger) for our method compared
to the standard method, since we know we are using a different model
to fit than to generate the data. However, comparing this under
various conditions can help us build intuition about when the proposed
method will perform well.
The second criterion we consider is the empirical coverage probability
of the 95\% prediction intervals, measured both across the range of
the input space and across repeated samples.

\begin{figure}

\includegraphics{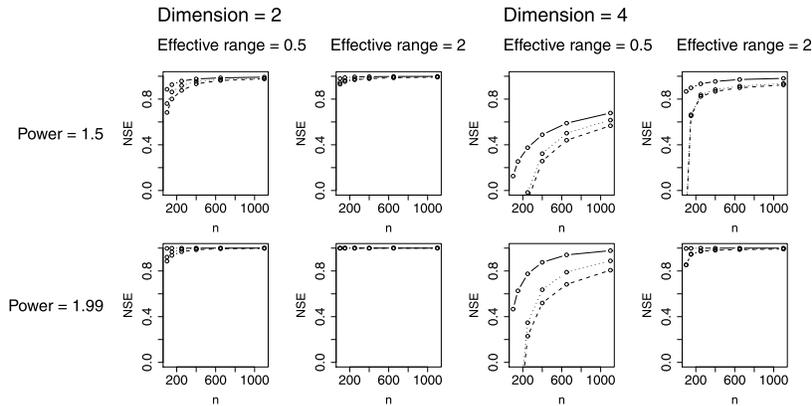}

\caption{Nash--Sutcliffe efficiencies for predictions made using the
posterior mean. In each panel, the solid lines corresponds to the
data-generating model, while the dashed and dotted lines correspond to
the proposed method with a maximum proportion of off-diagonal
elements of 0.02 or 0.05, respectively.}
\label{figCompareNSE}
\vspace*{2pt}
\end{figure}

We begin by comparing the NSE of the sparse method to that of the
standard method. Figure \ref{figCompareNSE} plots NSE under the eight
different combinations of dimension, power $\alpha$, and effective
%
%
range for which the data were generated. First examine the NSE for the
standard method, shown by the solid lines in each panel. It is clear
that the prediction task ranges quite a bit in difficulty across the
range of eight conditions, from processes that are difficult to
predict given the data (e.g., small NSE) to those which allow very
high accuracy predictions (with NSE close to~1). The prediction
problem is harder in four dimensions than in two, simply due to a
lower data density. Also, not surprisingly, the smoother and flatter
the process realizations, the easier they are to predict. That is,
holding other variables constant, the standard method has higher NSE
when the power is 1.99 rather than 1.5, and when the effective range
is 2.0 rather than 0.5.

Now examine the NSE for the sparse method. Not surprisingly, within
each panel the NSE is larger when the proportion of nonzero
off-diagonal elements is allowed to be 5\% (dotted line) rather than
only 2\% (dashed line). However, these differences are minor compared
to the differences across the different processes. Another interesting
trend that emerges is that, compared to the standard method, the
sparse method tends to perform quite well as the sample size gets
large. In most cases, by $n=1\mbox{,}100$ the sparse method is still capturing
about the same amount of the total variability as the standard method.
The proposed method does perform relatively worse when the
process is hard to predict, but even in these cases a large proportion
of the variability is explained. In light of the fact that we have not
expended any effort in determining the degree of the polynomial for
each realization of the GP, the results are even more encouraging. For
the large sample sizes for which the method was designed, we expect
the NSE to be close to 1.

\begin{figure}

\includegraphics{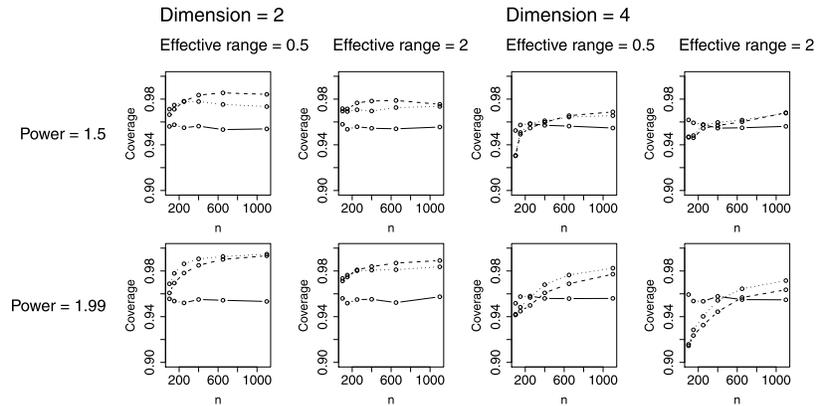}

\caption{Empirical coverage probabilities for pointwise credible
intervals. In each panel, the solid line corresponds to the
data-generating model, while the dashed and dotted lines correspond to
the proposed method with a maximum proportion of off-diagonal elements
of 0.02 or 0.05, respectively.}
\label{figCover}
\end{figure}

Next, consider the empirical coverage probabilities of the 95\%
prediction intervals. Figure \ref{figCover} shows the observed
%
%
coverage rates, under the same setup as in Figure~\ref{figCompareNSE}.
Not surprisingly, under the data-generating model, the coverage rates
are close to the nominal rate of 95\%. (The rates would be
theoretically exact if we were not also estimating the range
parameters.) Under the proposed model, the rates are often more
conservative than the nominal rate, particularly when the input has
only two dimensions. Although the widths of the intervals decrease with
$n$, they do not decrease rapidly enough to maintain only 95\%
coverage, a fact that is attributable to the shorter correlation ranges
being imposed for the sake of sparsity. This is a potential drawback to
using the proposed model, although we remind the reader that the
results under the data-generating model are not possible for large~$n$;
this is what motivates the approximation. These results should also not
be taken to be representative of all simulators one might encounter in
practice. In the application in Section \ref{example}, for example,
exploratory analysis with a~held-out sample suggests the coverage under
the proposed model is very close to the nominal rate. In that example,
posterior samples for the range parameters are well away from the
boundary imposed for sparsity.

\section{Application to photometric redshifts}\label{example}

A major aim of upcoming cosmological surveys is to characterize the
nature of dark energy, a mysterious type of energy that is driving a
current epoch of acceleration in the expansion of the Universe [for a
recent review, see \citet{Frieman2008}]. Evidence for cosmic
acceleration first came from measurements of the optical luminosity
from a specific class of supernovae [\citet{Riess1998},
\citet{Perlmutter1999}].
In a matter-dominated Universe, the expansion rate should slow down
with time, and the aim was to verify that the Universe was expanding
more rapidly in the past by studying distant supernovae. Instead,
observations indicated the Universe was doing exactly the opposite.
So puzzling is this behavior that understanding dark energy---a
hypothesized form of mass-energy accounting for the acceleration, or
perhaps signaling the breakdown of general relativity on very large
lengthscales---is considered to be one of the major unsolved
problems in all of physical science.

Information about dark energy can be inferred in a variety of ways,
some of which depend on an accurate three-dimensional representation
of galaxies in a cosmological survey. It is straightforward to
determine the angular location of an object in the sky, but the
fundamental difficulty is determining the distance with sufficient
accuracy. In large scale structure studies, the analogue of radial
distance is the cosmological redshift. Due to cosmic expansion, the
wavelength of light received from a distant object is stretched
(``redshifted''), with more distant objects being at greater
redshifts. Accurate redshift determination requires measurement of the
spectrum of each galaxy at high resolution, but this is very difficult
and too time-consuming for the very large numbers of very distant,
hence very dim, galaxies. An alternative is to obtain galaxy
photometry (flux measurements) in a restricted number of wavebands,
and to attempt to reconstruct the redshift from just this information.
Redshifts obtained in this way are termed ``photometric'' redshifts,
in contrast to the more accurate ``spectroscopic'' redshifts. To
obtain a photometric redshift estimate, however, requires estimating
the functional relationship between the observations within the
wavebands and the spectroscopic redshift.

The simulation we
consider here models this relationship for four different wavebands.
It simulates the true spectroscopic redshift and corresponding
photometric measurements for the Dark Energy Survey
[\citet{Abbott2005}, \citet{Oyaizu2006}], which will come
online in the near
future. The training and validation data sets were generated to be of
size 20,000 and
80,000, respectively. The design points were not chosen systematically,
but were rather sampled from distributions meant to mimic what will be
encountered in data from the Dark Energy Survey. The analysis in this
article treats the simulation as the usual noiseless computer model
case, with inputs corresponding to the flux measurements in the four
wavebands and output corresponding to the spectroscopic redshift. There
is some expected intrinsic scatter due to
coarsening of information, potential degeneracies, and the fact that
the predictors can
be viewed as often having error themselves. Nonetheless, we expect that
the GP will
be able to efficiently predict the spectroscopic redshift from model inputs.

Our analysis proceeds in two steps. We first carry out an exploratory
analysis on a small subset of the training data, so that we may
explore the effects of various modeling choices before implementing
these choices on the entire data set. Interestingly, we show here that
the methods outlined in this paper in fact outperform the standard
model in both predictive as well as computational efficiency. That is,
the computationally efficient methods that allow us to scale up and
work with the full data set do not come at a cost to predictive
efficiency; they actually have \textit{higher} Nash--Sutcliffe efficiency
than the traditional, computationally infeasible methods. This gives
us confidence in proceeding to the second step of the analysis, which
is to fit the model on the set of 20,000 training points and evaluate
the predictions at the 80,000 validation points.\vspace*{-1pt}

\subsection{Preliminary analysis and model comparison}\label{sec51}
We first normalize the input space so that each input variable lies
within $[0,1]$. We sample a smaller ``training'' and ``validation''
set from the full training set of 20,000; these have size 2,000 and
500, respectively, which is small enough to fit the standard model for
comparison.

\begin{figure}

\includegraphics{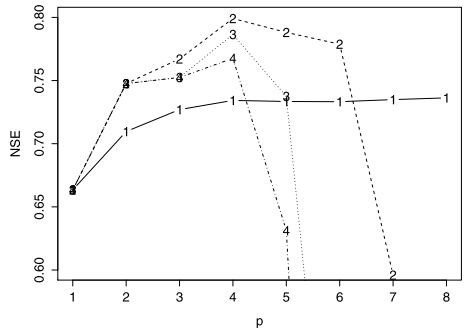}

\caption{Nash--Sutcliffe efficiency (NSE) of predictions made using
OLS estimates and various choices of maximum degree $p$ and
maximum interacting dimensions $m$ (indicated by number on the plot).
The choice $p=4$ and $m=2$
produces the highest NSE and includes relatively few regression
terms ($q=53$) compared to the sample size.}
\label{figdegreetest}
\end{figure}

The first question to be addressed in this exploratory analysis is the
choice of basis functions $f_i$ in (\ref{surface}). As in the
simulation study, we consider various tensor products of Legendre
polynomials over $[0,1]$. Let $p$ represent the maximum degree, that is,
the maximum power in a main effect for a single input dimension, or
the maximum sum of powers in an interaction. Let $m$ represent the
maximum number of dimensions that may be involved in an interaction.
Then\vspace*{1.5pt} a simple way to evaluate the choice of $p$ and $m$ is to find the
ordinary least squares estimates $\hat\beta$ for each choice using
the $n=2\mbox{,}000$ training set and evaluate the Nash--Sutcliffe efficiency
of predictions~$X_0\hat\beta$ for~$X_0$ consisting of the
corresponding regression terms evaluated at the validation input set
with $n=500$. This is shown in Figure~\ref{figdegreetest}. The choice
$p=4$ and $m=2$ produces the largest NSE, of 0.799, and produces
relatively few regression terms ($q=53$) compared to the sample size:
here 2,000, later 20,000. Although matrix multiplication requires many
fewer evaluations than the solution of a~linear system of equivalent
size, it can still be problematic if both $q$ and~$n$ are large, so it
is a happy coincidence that the choice of $p$ and $m$ that has highest
NSE for this data is also computationally efficient.

%

We now fit the model (\ref{surface}) with all combinations of the
following choices:
\begin{itemize}
\item the covariance structure $R$ in (\ref{covfunction}) set to
either the power exponential function (\ref{powerexp}) or a product
of truncated power functions (\ref{truncpow}) in each dimension,
\item the parameter $\alpha$ in (\ref{powerexp}) and (\ref{truncpow})
set either to 1, $3/2$ or $5/3$ in each dimension,
\item the mean including an intercept only ($q=1$) or a regression on
tensor products of Legendre polynomials as described above, with
$p=4$ and $m=2$.
\end{itemize}

In all models, we used the prior specification $p(\sigma^2,\beta)
\propto1/\sigma^2$. We took the prior for $\bolds\phi$ in
(\ref{powerexp}) to be uniform over a hyper-cube, and, as specified in
(\ref{triangle}), we took the prior for $\tau$ to be uniform over a
hyper-triangle. In particular, we took the cutoff $C=0.185$, chosen so
that the proportion of off-diagonal elements in the correlation matrix
that were nonzero would be at most 2\%. This imposes a very high
degree of sparsity, and in this initial exploratory analysis we can
determine the effect of this choice on the posterior distribution for
$\bolds\tau$.

For each of the twelve modeling combinations, we sampled from the
posterior distribution for model parameters and used this to generate
predictions and pointwise credible intervals for the validation set.
Table \ref{nsetable} shows the Nash--Sutcliffe efficiency for each
modeling combination, and Table \ref{cvgtable} shows the empirical
coverage probabilities. First note that the largest NSE in Table
\ref{nsetable} is for the sparse correlation structure
(\ref{truncpow}) with power $\alpha= 1$ and Legendre polynomials up
to degree 4, followed closely by the same model with $\alpha= 3/2$.
This relationship holds across all entries in rows three and
four of the table, corresponding to a model with $q=53$ regression
terms. It is interesting to note that this trend reverses in the first
two rows, corresponding to the intercept only model. This provides
evidence that in our method, both components---compactly supported
correlation structure as well as a more structured mean term---are
needed to achieve good predictive accuracy. Table \ref{cvgtable}
shows that the empirical coverage of the prediction intervals is very close
to the nominal 95\% level when using regression terms and a sparse
correlation. This
again gives us confidence in our method.

%
\begin{table}
\tablewidth=300pt
\caption{Nash--Sutcliffe efficiencies for predictions made using the
posterior mean}\label{nsetable}
\begin{tabular*}{\tablewidth}{@{\extracolsep{\fill}}ccccc@{}}
\hline
& & $\bolds{\mathrm{Power}=1}$ & $\bolds{\mathrm{Power}=3/2}$
& $\bolds{\mathrm{Power}=5/3}$ \\
\hline
$\mathrm{Degree} = 0$ & Nonsparse & 0.791 & 0.773 & 0.757 \\
& Sparse & 0.702 & 0.659 & 0.617 \\
[4pt]
$\mathrm{Degree} = 4$ & Nonsparse & 0.839 & 0.818 & 0.761 \\
& Sparse & 0.849 & 0.848 & 0.843 \\
\hline
\end{tabular*}
\end{table}

%
\begin{table}[b]
\tablewidth=300pt
\caption{Empirical coverage probabilities for posterior predictive
intervals}\label{cvgtable}
\begin{tabular*}{\tablewidth}{@{\extracolsep{\fill}}ccccc@{}}
\hline
& & $\bolds{\mathrm{Power} = 1}$ & $\bolds{\mathrm{Power} = 3/2}$
& $\bolds{\mathrm{Power} = 5/3}$ \\
\hline
$\mathrm{Degree} = 0$ & Nonsparse & 0.936 & 0.928 & 0.908 \\
& Sparse & 0.960 & 0.950 & 0.948 \\
[4pt]
$\mathrm{Degree} = 4$ & Nonsparse & 0.952 & 0.926 & 0.906 \\
& Sparse & 0.954 & 0.952 & 0.952 \\
\hline
\end{tabular*}
\end{table}

\begin{figure}

\includegraphics{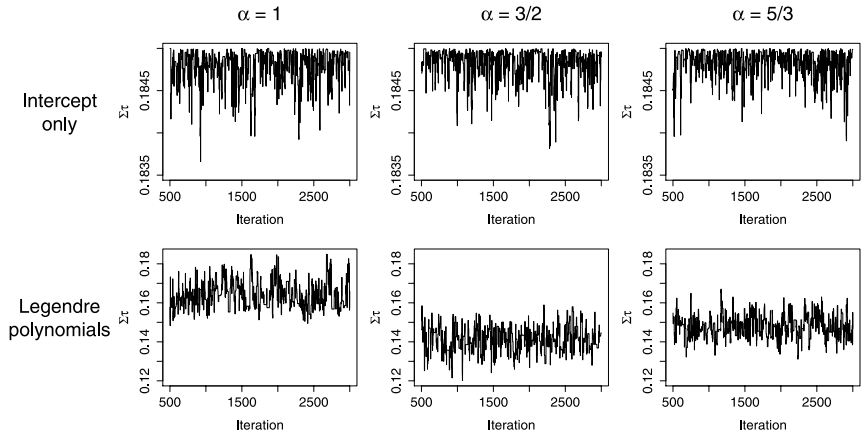}

\caption{Trace plots\vspace*{2pt} for $\sum_{j=1}^4\tau_j$ over iterations in
the Metropolis sampler. Introducing regression terms into the
mean of the Gaussian process (degree${}={}$4) produces much better
mixing, as well as the potential for increased computational
efficiency.}
\label{figtautrace}
\end{figure}

There is one final comparison we make using this
preliminary test set. This is for the sparse methods only,
corresponding to rows two and four of the tables. Note that the prior
distribution restricts $\sum_{j=1}^4 \tau_j \leq C$, where we chose
$C=0.185$. Figure \ref{figtautrace} shows the trace plots of this sum
over iterations of the Metropolis sampler, discarding an initial
%
%
burn-in period. Note another major
implication of using the Legendre polynomials: it changes the
posterior distribution for $\bolds\tau$, so that the posterior
distribution of the sum moves away from this boundary. In contrast, in
the intercept only model, the posterior samples of $\bolds\tau$
are varying only slightly around the upper boundary of 0.185. (Note the
difference in scales between rows in Figure~\ref{figtautrace}.)
Another way of saying this is that a cutoff of $C=0.185$ is too small
for the intercept only GP model to capture all the variability in the
data set, but it is adequate to capture variability in the residuals
after introducing the Legendre polynomials.
This has
implications both for mixing of the sampler, which is much better in
the models with the regression terms, and for computational efficiency,
as we see that
the cutoff $C$ can be reduced even further, as far as 0.16 when the
power $\alpha=3/2$. The smaller this cutoff, the larger the
computational savings, as it allows us to rule out more pairs of input
values before the MCMC even begins to run. For this reason, as well as
observing that the Nash--Sutcliffe efficiency when $\alpha=3/2$ was
only very slightly larger than the optimal one in Table
\ref{nsetable}, we choose to use this setting when working with the
full data set.

\subsection{Full analysis}

\begin{figure}

\includegraphics{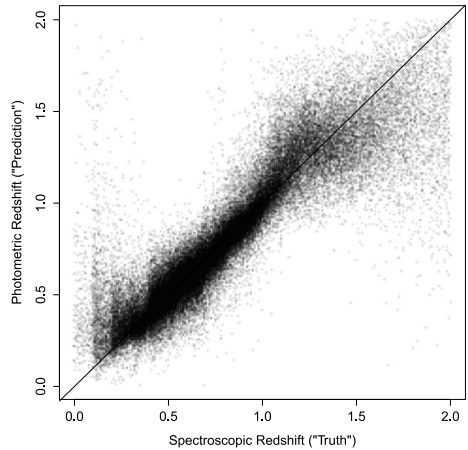}

\caption{Photometric redshift predictions and the corresponding
spectroscopic redshift values from the simulator. Points are
plotted with transparency, so that dark areas of the plot
indicate a high density of points being over-plotted in the same
region.}
\label{figpredictions}
\end{figure}

In the second stage of the analysis, we fit the model using the full
set of $20\mbox{,}000$ training points and made predictions at each of
the $80\mbox{,}000$ validation points. We sampled $B=3\mbox{,}000$ MCMC
iterations using the Gibbs sampling algorithm described in Section
\ref{implementation} and the \hyperref[app]{Appendix}, discarding the
first 500 for burn-in. For every tenth iteration thereafter, we
calculated the conditional means and variances of the predictive
distribution for $Y(x_0)$ given the observations and the parameter
values at that iteration, from which we formed Monte Carlo
approximations of the mean and variance of the posterior predictive
distribution, as also described in the \hyperref [app]{Appendix}. The
posterior means for the 80,000 new input values are shown in Figure
\ref{figpredictions}, plotted against the actual spectroscopic redshift
%
%
values. The Nash--Sutcliffe efficiency for the predictions is 0.831, and
the empirical coverage rate for the 95\% credible intervals is close to
the nominal level, at 94\%.

\section{Discussion}\label{sec6}

In this article we have proposed new methodology for analyzing large
computer experiments using a GP. The approach uses an
anisotropic\vadjust{\goodbreak}
compactly supported covariance, as well as a regression function for
mean, to emulate the computer model. With respect to the latter
adaptation, we have proposed the use of a flat prior distribution for
the regression parameters $\bolds\beta$, using a preliminary
study to select the set of basis functions to be used. This provides
additional computational efficiency, as all parameters except
$\bolds\tau$ may be integrated out of the model. Given the model
selection problem, which we solve in a rather ad-hoc way, one might
also be tempted to incorporate a model selection or model averaging
approach. For example, one approach we examined was to err on the side
of including more basis functions, but to use a shrinkage prior for
$\bolds\beta$, with
%
%
\begin{eqnarray}\label{ridge}
\beta_i | \xi_i &\stackrel{\mathrm{indep}}{\sim}& N(0, \xi_i),\qquad i =
1,\ldots,p, \nonumber\\[-8pt]\\[-8pt]
\xi_i &\stackrel{\mathrm{i.i.d.}}{\sim}& IG(a,b),\qquad i = 1,\ldots,p
\nonumber.
\end{eqnarray}
The prior specification in (\ref{ridge}) is akin to a generalized
ridge regression, in which each coefficient receives its own shrinkage
weight $w_i = \xi_i/(1+\xi_i)$ [\citet{Dension2000}].
Despite the elegance of this approach, we found that it contributed
very little to the predictive efficiency of our method; it increased
the Nash--Sutcliffe efficiency by only a few percentage points, and
then only for small data sets, not the type that motivate our work. In
our judgement, the added computational cost
is not worth this small
potential improvement. Instead, we advocate the simpler and
computationally more efficient approach of choosing the basis
functions based on a random subset of the data in a preliminary study,
as described in Section \ref{example}.

We also note that the inclusion of the regression terms in the mean of
the GP has implications for extrapolation beyond the range of the
initial input values. Although polynomial regression can be problematic
when it comes to extrapolation, it is unclear that the behavior
obtained under a constant mean term is more desirable. In the standard
model, when one is far from the initial input values, the GP prediction
returns to the global mean and does not depend on the new inputs at
all. Indeed, \citet{Bayarri2007} suggest using a more complicated
mean structure to avoid this problem. The fact that neither approach is
entirely satisfactory reflects the difficulty of extrapolation, and
users should be aware of the implications of the structure of the mean
term if it must be undertaken.

\begin{appendix}\label{app}
\section*{Appendix: Posterior sampling and prediction}

To generate samples from the posterior distribution $p(\bolds
\tau| \mathbf Y)$, we use an adaptive Metropolis algorithm, taking the
transition density to be a multivariate normal random walk. At
iteration $i$, we sample a candidate $ \bolds\tau^{\mathrm{cand}}
|\allowbreak
\bolds\tau^{(i-1)} \sim \mathit{MVN}( \bolds\tau^{(i-1)},
\Sigma^{(i)})$, where $\Sigma^{(i)}$ is calculated adaptively using an
algorithm to be described shortly. If the candidate value falls
outside of the constrained parameter space $T_C$, as defined in
(\ref{triangle}), it is immediately rejected and we set
$\bolds\tau^{(i)}=\bolds\tau^{(i-1)}$. Otherwise, we
calculate the integrated likelihood~$L^I(\bolds\tau^{\mathrm{cand}})$,
where
\[
L^I( {\bolds\tau}) \propto|\bolds\Gamma({\bolds
\tau})|^{-1/2} |\mathbf{F}'\bolds\Gamma({\bolds\tau})^{-1}\mathbf{F}|^{-1/2}(\hat\sigma^2
({\bolds
\tau}))^{(n-p)/2}
\]
for $\hat\sigma^2({\bolds\tau}) = (\mathbf Y-\mathbf{F}{\hat\beta}
({\bolds
\tau}))' \bolds\Gamma({\bolds\tau})^{-1} (\mathbf
Y-\mathbf{F}\hat\beta({\bolds\tau}))$ and $\hat\beta({\bolds\tau})
= (\mathbf F'\bolds\Gamma({\bolds\tau})^{-1}\mathbf{F})^{-1}
\mathbf{F}'\times\bolds\Gamma({\bolds\tau})^{-1} \mathbf Y$. The computationally
demanding aspects of this calculation are described in Section
\ref{implementation}. We then set
$\bolds\tau^{(i)} = \bolds\tau^{\mathrm{cand}}$ with probability\break
$\max\{L^I(\bolds\tau^{\mathrm{cand}})/ L^I(\bolds\tau^{(i-1)}),1\}$
and $\bolds\tau^{(i-1)}$ otherwise.

We adapt the proposal covariance
matrix using the Log-Adaptive Proposal algorithm of \citet{Shaby2011},
a slightly modified version of Algorithm 4 in
\citet{Andrieu2008}. The algorithm periodically updates an
estimate of
the posterior covariance matrix and then takes the proposal covariance
matrix to be a scaled version of this, the scaling factor also being
updated. This allows the sampler to target a particular
acceptance rate and thereby increase efficiency. Although using
previous states to generate the proposal violates the Markov property
of the chain, the ergodicity of the process is maintained within a
framework of ``vanishing adaptation''
[\citet{Roberts2008}].

After sampling from the posterior distribution for $\bolds\tau$,
the second step is to generate samples from the posterior predictive
distribution for the output of the simulator at new input values. Let
$\mathbf Y_0$ denote this new output. One could generate a sample from
$p(\mathbf{Y}_0 | \mathbf{Y}, \bolds\tau)$ at each iteration of the
MCMC. However, since we are primarily
interested in the mean and variance of the predictive distribution, we
instead adopt the computationally more stable approach of calculating
conditional means and variances at each iteration (i.e.,
conditional on $\bolds\tau^{(i)}$) and using these to reconstruct
the predictive mean and variance. Specifically, for a subset of the
$\bolds\tau$
samples (the number of which may be chosen based on an estimate of the
smallest effective sample size among the elements of $\bolds\tau
$), we calculate the mean and variance of $\mathbf
Y_0$ given $\mathbf Y$ and $\bolds\tau^{(i)}$,
to produce $\mathbf{m}^{(i)}$ and $\mathbf{v}^{(i)}$. As
\[
\pmatrix{ \mathbf{Y} \cr \mathbf{Y}_0 }
\big| \sigma^2, \bolds\beta, \bolds\tau\sim \mathit{MVN}
\left(
\pmatrix{ \mathbf X \cr\mathbf X_0 } \bolds\beta,
\sigma^2 \pmatrix{
\bolds\Gamma(\bolds\tau) & \bolds\gamma(\bolds\tau)\cr
\bolds\gamma'(\bolds\tau) & \bolds\Gamma_0(\bolds\tau)}
\right),
\]
$p(\mathbf{Y}_0|\mathbf{Y}, \bolds\tau^{(i)})$ is multivariate t, with
$\mathbf{m}^{(i)}\,{=}\,\hat{Y}_0(\bolds\tau^{(i)})$ and
$\mathbf{v}^{(i)}\,{=}\,\frac{n-q}{n-q-2}
\hat\sigma^2(\bolds\tau^{(i)})\times
\mathbf{V}(\bolds\tau^{(i)})$,
for
\begin{eqnarray*}
\hat{\mathbf{Y}}_0(\bolds\tau) &= & \mathbf{X}_0 \hat\beta(\bolds\tau
) + \bolds\gamma'(\bolds\tau)\bolds\Gamma(\bolds\tau)^{-1}[\mathbf{Y} -
\mathbf{X}\hat\beta(\bolds\tau)],\\
\hat\beta(\bolds\tau) &=& \bigl(\mathbf X'\bolds\Gamma(\bolds\tau
)^{-1}\mathbf{X}\bigr)^{-1}\mathbf{X}'\bolds\Gamma(\bolds\tau)^{-1}\mathbf{Y},\\
\hat\sigma^2(\bolds\tau) &=& \bigl(\mathbf Y-\mathbf{X}\hat\beta(\bolds\tau
)\bigr)'\bolds\Gamma(\bolds\tau)^{-1}\bigl(\mathbf{Y}-\mathbf{X}\hat\beta(\bolds\tau)\bigr)/\mathbf{n},\\
V(\bolds\tau) &=& \bolds\Gamma_0(\bolds\tau) - \bolds\gamma
(\bolds\tau)'\bolds\Gamma(\bolds\tau)^{-1}\bolds\gamma
(\bolds\tau) + \mathbf{X}_0\bigl(\mathbf{X}'\bolds\Gamma(\bolds\tau)^{-1}\mathbf{X}\bigr)^{-1}\mathbf{X}_0.
\end{eqnarray*}

Finally, we use Monte Carlo approximation to estimate $E[\mathbf Y(\mathbf{x}_0)|\mathbf{Y}]$ and $\operatorname{Var}
[\mathbf Y(\mathbf{x}_0)|\mathbf{Y}]$ from the $K$ samples
according to
\begin{eqnarray*}
E[\mathbf Y(\mathbf{x}_0)|\mathbf{Y}]
&=& E[E[\mathbf Y(\mathbf{x}_0)|\mathbf{Y}, \bolds\tau]] \\
&\approx&\frac{1}{K} \sum_{i=1}^K \mathbf{m}^{(i)} \equiv\bar{\mathbf
m},
\\
\operatorname{Var}[\mathbf Y(\mathbf{x}_0)|\mathbf{Y}]
&=& E[\operatorname{Var}[\mathbf Y(\mathbf{x}_0)|\mathbf{Y}, \bolds\tau]] + \operatorname{Var}[E
[\mathbf Y(\mathbf{x}_0)|\mathbf{Y}, \bolds\tau]]\\
&\approx&\frac{1}{K} \sum_{i=1}^K \mathbf{v}^{(i)} + \frac{1}{K}\sum
_{i=1}^K\bigl(\mathbf{m}^{(i)} - \bar{\mathbf m}\bigr)^2.
\end{eqnarray*}
These can be used to construct approximate pointwise credible
intervals for~$Y(\mathbf{x}_0)$ given $\mathbf{Y}$. If global intervals are
needed, one should instead sample from the corresponding multivariate $t$
conditional distributions at each iteration.
\end{appendix}

\section*{Acknowledgments}

Habib and Heitmann acknowledge support from the LDRD program at Los
Alamos National Laboratory. Habib and Heitmann also acknowledge the
Aspen Center for Physics, where part of this work was initialized. We
thank the Associate Editor and an anonymous reviewer for their helpful
comments on the manuscript.



%
\printaddresses


\begin{thebibliography}{35}

\bibitem[\protect\citeauthoryear{Abbott et~al.}{2005}]{Abbott2005}
\begin{bmisc}[author]
\bauthor{\bsnm{Abbott},~\bfnm{T}\binits{T.}} \betal{et~al.}
(\byear{2005}).
\bhowpublished{The dark energy survey.
Preprint. Available at
\href{http://xxx.lanl.gov/archive/astro-ph/0510346}{Astro-ph/}
\href{http://xxx.lanl.gov/archive/astro-ph/0510346}{0510346}.}
\bptok{imsref}%
\end{bmisc}
\endbibitem

\bibitem[\protect\citeauthoryear{An and Owen}{2001}]{An2001}
\begin{barticle}[mr]
\bauthor{\bsnm{An},~\bfnm{Jian}\binits{J.}} \AND
  \bauthor{\bsnm{Owen},~\bfnm{Art}\binits{A.}}
(\byear{2001}).
\btitle{Quasi-regression}.
\bjournal{J. Complexity}
\bvolume{17}
\bpages{588--607}.
\bid{doi={10.1006/jcom.2001.0588}, issn={0885-064X}, mr={1881660}}
\bptok{imsref}%
\end{barticle}
\endbibitem

\bibitem[\protect\citeauthoryear{Andrieu and Thoms}{2008}]{Andrieu2008}
\begin{barticle}[mr]
\bauthor{\bsnm{Andrieu},~\bfnm{Christophe}\binits{C.}} \AND
  \bauthor{\bsnm{Thoms},~\bfnm{Johannes}\binits{J.}}
(\byear{2008}).
\btitle{A tutorial on adaptive {MCMC}}.
\bjournal{Stat. Comput.}
\bvolume{18}
\bpages{343--373}.
\bid{doi={10.1007/s11222-008-9110-y}, issn={0960-3174}, mr={2461882}}
\bptok{imsref}%
\end{barticle}
\endbibitem

\bibitem[\protect\citeauthoryear{Barry and Pace}{1997}]{Barry1997}
\begin{barticle}[author]
\bauthor{\bsnm{Barry},~\bfnm{R.~P.}\binits{R.~P.}} \AND
  \bauthor{\bsnm{Pace},~\bfnm{R.~K.}\binits{R.~K.}}
(\byear{1997}).
\btitle{Kriging with large data sets using sparse matrix techniques}.
\bjournal{Comm. Statist. Simulation Comput.}
\bvolume{26}
\bpages{619--629}.
\bptok{imsref}%
\end{barticle}
\endbibitem

\bibitem[\protect\citeauthoryear{Bayarri et~al.}{2007}]{Bayarri2007}
\begin{barticle}[mr]
\bauthor{\bsnm{Bayarri},~\bfnm{Maria~J.}\binits{M.~J.}},
  \bauthor{\bsnm{Berger},~\bfnm{James~O.}\binits{J.~O.}},
  \bauthor{\bsnm{Paulo},~\bfnm{Rui}\binits{R.}},
  \bauthor{\bsnm{Sacks},~\bfnm{Jerry}\binits{J.}},
  \bauthor{\bsnm{Cafeo},~\bfnm{John~A.}\binits{J.~A.}},
  \bauthor{\bsnm{Cavendish},~\bfnm{James}\binits{J.}},
  \bauthor{\bsnm{Lin},~\bfnm{Chin-Hsu}\binits{C.-H.}} \AND
  \bauthor{\bsnm{Tu},~\bfnm{Jian}\binits{J.}}
(\byear{2007}).
\btitle{A framework for validation of computer models}.
\bjournal{Technometrics}
\bvolume{49}
\bpages{138--154}.
\bid{doi={10.1198/004017007000000092}, issn={0040-1706}, mr={2380530}}
\bptok{imsref}%
\end{barticle}
\endbibitem

\bibitem[\protect\citeauthoryear{Berger, De~Oliveira and
  Sans{\'o}}{2001}]{Berger2001}
\begin{barticle}[mr]
\bauthor{\bsnm{Berger},~\bfnm{James~O.}\binits{J.~O.}},
  \bauthor{\bsnm{De~Oliveira},~\bfnm{Victor}\binits{V.}} \AND
  \bauthor{\bsnm{Sans{\'o}},~\bfnm{Bruno}\binits{B.}}
(\byear{2001}).
\btitle{Objective {B}ayesian analysis of spatially correlated data}.
\bjournal{J. Amer. Statist. Assoc.}
\bvolume{96}
\bpages{1361--1374}.
\bid{doi={10.1198/016214501753382282}, issn={0162-1459}, mr={1946582}}
\bptok{imsref}%
\end{barticle}
\endbibitem

\bibitem[\protect\citeauthoryear{Cressie}{1993}]{Cressie1993}
\begin{bbook}[mr]
\bauthor{\bsnm{Cressie},~\bfnm{Noel A.~C.}\binits{N.~A.~C.}}
(\byear{1993}).
\btitle{Statistics for Spatial Data}.
\bpublisher{Wiley}, \baddress{New York}.
\bid{mr={1239641}}
\bptok{imsref}%
\end{bbook}
\endbibitem

\bibitem[\protect\citeauthoryear{Denison and George}{2000}]{Dension2000}
\begin{bmisc}[author]
\bauthor{\bsnm{Denison},~\bfnm{D.}\binits{D.}} \AND
  \bauthor{\bsnm{George},~\bfnm{E.}\binits{E.}}
(\byear{2000}).
\bhowpublished{Bayesian prediction using adaptive ridge estimators.
Technical report, Dept. Mathematics, Imperial College, London, UK.}
\bptok{imsref}%
\end{bmisc}
\endbibitem

\bibitem[\protect\citeauthoryear{Frieman, Turner and
  Huterer}{2008}]{Frieman2008}
\begin{barticle}[author]
\bauthor{\bsnm{Frieman},~\bfnm{J.~A.}\binits{J.~A.}},
  \bauthor{\bsnm{Turner},~\bfnm{M.~S.}\binits{M.~S.}} \AND
  \bauthor{\bsnm{Huterer},~\bfnm{D.}\binits{D.}}
(\byear{2008}).
\btitle{Dark energy and the accelerating universe}.
\bjournal{Annual Review of Astronomy and Astrophysics}
\bvolume{46}
\bpages{385--432}.
\bptok{imsref}%
\end{barticle}
\endbibitem

\bibitem[\protect\citeauthoryear{Furrer, Genton and Nychka}{2006}]{Furrer2006}
\begin{barticle}[mr]
\bauthor{\bsnm{Furrer},~\bfnm{Reinhard}\binits{R.}},
  \bauthor{\bsnm{Genton},~\bfnm{Marc~G.}\binits{M.~G.}} \AND
  \bauthor{\bsnm{Nychka},~\bfnm{Douglas}\binits{D.}}
(\byear{2006}).
\btitle{Covariance tapering for interpolation of large spatial datasets}.
\bjournal{J. Comput. Graph. Statist.}
\bvolume{15}
\bpages{502--523}.
\bid{doi={10.1198/106186006X132178}, issn={1061-8600}, mr={2291261}}
\bptok{imsref}%
\end{barticle}
\endbibitem

\bibitem[\protect\citeauthoryear{Furrer and Sain}{2010}]{Furrer2010}
\begin{barticle}[author]
\bauthor{\bsnm{Furrer},~\bfnm{Reinhard}\binits{R.}} \AND
  \bauthor{\bsnm{Sain},~\bfnm{Stephan~R.}\binits{S.~R.}}
(\byear{2010}).
\btitle{{spam}: A sparse matrix {R} package with emphasis on MCMC methods for
  Gaussian Markov random fields}.
\bjournal{Journal of Statistical Software}
\bvolume{36}
\bpages{1--25}.
\bptok{imsref}%
\end{barticle}
\endbibitem

\bibitem[\protect\citeauthoryear{Gneiting}{2001}]{Gneiting2001}
\begin{barticle}[mr]
\bauthor{\bsnm{Gneiting},~\bfnm{Tilmann}\binits{T.}}
(\byear{2001}).
\btitle{Criteria of {P}\'olya type for radial positive definite functions}.
\bjournal{Proc. Amer. Math. Soc.}
\bvolume{129}
\bpages{2309--2318 (electronic)}.
\bid{doi={10.1090/S0002-9939-01-05839-7}, issn={0002-9939}, mr={1823914}}
\bptok{imsref}%
\end{barticle}
\endbibitem

\bibitem[\protect\citeauthoryear{Gneiting}{2002}]{Gneiting2002}
\begin{barticle}[mr]
\bauthor{\bsnm{Gneiting},~\bfnm{Tilmann}\binits{T.}}
(\byear{2002}).
\btitle{Compactly supported correlation functions}.
\bjournal{J. Multivariate Anal.}
\bvolume{83}
\bpages{493--508}.
\bid{doi={10.1006/jmva.2001.2056}, issn={0047-259X}, mr={1945966}}
\bptok{imsref}%
\end{barticle}
\endbibitem

\bibitem[\protect\citeauthoryear{Golubov}{1981}]{Golubov1981}
\begin{barticle}[mr]
\bauthor{\bsnm{Golubov},~\bfnm{B.~I.}\binits{B.~I.}}
(\byear{1981}).
\btitle{On {A}bel--{P}oisson type and {R}iesz means}.
\bjournal{Anal. Math.}
\bvolume{7}
\bpages{161--184}.
\bid{doi={10.1007/BF01908520}, issn={0133-3852}, mr={0635483}}
\bptok{imsref}%
\end{barticle}
\endbibitem

\bibitem[\protect\citeauthoryear{Irvine, Gitelman and
  Hoeting}{2007}]{Irvine2007}
\begin{barticle}[mr]
\bauthor{\bsnm{Irvine},~\bfnm{Kathryn~M.}\binits{K.~M.}},
  \bauthor{\bsnm{Gitelman},~\bfnm{Alix~I.}\binits{A.~I.}} \AND
  \bauthor{\bsnm{Hoeting},~\bfnm{Jennifer~A.}\binits{J.~A.}}
(\byear{2007}).
\btitle{Spatial designs and properties of spatial correlation: Effects on
  covariance estimation}.
\bjournal{J. Agric. Biol. Environ. Stat.}
\bvolume{12}
\bpages{450--469}.
\bid{doi={10.1198/108571107X249799}, issn={1085-7117}, mr={2405534}}
\bptok{imsref}%
\end{barticle}
\endbibitem

\bibitem[\protect\citeauthoryear{Kaufman, Schervish and
  Nychka}{2008}]{Kaufman2008}
\begin{barticle}[mr]
\bauthor{\bsnm{Kaufman},~\bfnm{Cari~G.}\binits{C.~G.}},
  \bauthor{\bsnm{Schervish},~\bfnm{Mark~J.}\binits{M.~J.}} \AND
  \bauthor{\bsnm{Nychka},~\bfnm{Douglas~W.}\binits{D.~W.}}
(\byear{2008}).
\btitle{Covariance tapering for likelihood-based estimation in large spatial
  data sets}.
\bjournal{J. Amer. Statist. Assoc.}
\bvolume{103}
\bpages{1545--1555}.
\bid{doi={10.1198/016214508000000959}, issn={0162-1459}, mr={2504203}}
\bptok{imsref}%
\end{barticle}
\endbibitem

\bibitem[\protect\citeauthoryear{Kennedy and O'Hagan}{2001}]{Kennedy2001}
\begin{barticle}[mr]
\bauthor{\bsnm{Kennedy},~\bfnm{Marc~C.}\binits{M.~C.}} \AND
  \bauthor{\bsnm{O'Hagan},~\bfnm{Anthony}\binits{A.}}
(\byear{2001}).
\btitle{Bayesian calibration of computer models}.
\bjournal{J.~R.~Stat. Soc. Ser. B Stat. Methodol.}
\bvolume{63}
\bpages{425--464}.
\bid{doi={10.1111/1467-9868.00294}, issn={1369-7412}, mr={1858398}}
\bptok{imsref}%
\end{barticle}
\endbibitem

\bibitem[\protect\citeauthoryear{Linkletter et~al.}{2006}]{Linkletter2006}
\begin{barticle}[mr]
\bauthor{\bsnm{Linkletter},~\bfnm{Crystal}\binits{C.}},
  \bauthor{\bsnm{Bingham},~\bfnm{Derek}\binits{D.}},
  \bauthor{\bsnm{Hengartner},~\bfnm{Nicholas}\binits{N.}},
  \bauthor{\bsnm{Higdon},~\bfnm{David}\binits{D.}} \AND
  \bauthor{\bsnm{Ye},~\bfnm{Kenny~Q.}\binits{K.~Q.}}
(\byear{2006}).
\btitle{Variable selection for {G}aussian process models in computer
  experiments}.
\bjournal{Technometrics}
\bvolume{48}
\bpages{478--490}.
\bid{doi={10.1198/004017006000000228}, issn={0040-1706}, mr={2328617}}
\bptok{imsref}%
\end{barticle}
\endbibitem

\bibitem[\protect\citeauthoryear{McKay, Beckman and Conover}{1979}]{McKay1979}
\begin{barticle}[mr]
\bauthor{\bsnm{McKay},~\bfnm{M.~D.}\binits{M.~D.}},
  \bauthor{\bsnm{Beckman},~\bfnm{R.~J.}\binits{R.~J.}} \AND
  \bauthor{\bsnm{Conover},~\bfnm{W.~J.}\binits{W.~J.}}
(\byear{1979}).
\btitle{A comparison of three methods for selecting values of input variables
  in the analysis of output from a computer code}.
\bjournal{Technometrics}
\bvolume{21}
\bpages{239--245}.
\bid{issn={0040-1706}, mr={0533252}}
\bptok{imsref}%
\end{barticle}
\endbibitem

\bibitem[\protect\citeauthoryear{Oakley and O'Hagan}{2004}]{Oakley2004}
\begin{barticle}[mr]
\bauthor{\bsnm{Oakley},~\bfnm{Jeremy~E.}\binits{J.~E.}} \AND
  \bauthor{\bsnm{O'Hagan},~\bfnm{Anthony}\binits{A.}}
(\byear{2004}).
\btitle{Probabilistic sensitivity analysis of complex models: A~{B}ayesian
  approach}.
\bjournal{J. R. Stat. Soc. Ser. B Stat. Methodol.}
\bvolume{66}
\bpages{751--769}.
\bid{doi={10.1111/j.1467-9868.2004.05304.x}, issn={1369-7412}, mr={2088780}}
\bptok{imsref}%
\end{barticle}
\endbibitem

\bibitem[\protect\citeauthoryear{Oyaizu et~al.}{2006}]{Oyaizu2006}
\begin{binproceedings}[author]
\bauthor{\bsnm{Oyaizu},~\bfnm{H.}\binits{H.}},
  \bauthor{\bsnm{Cunha},~\bfnm{C.}\binits{C.}},
  \bauthor{\bsnm{Lima},~\bfnm{M.}\binits{M.}},
  \bauthor{\bsnm{Lin},~\bfnm{H.}\binits{H.}} \AND
  \bauthor{\bsnm{Frieman},~\bfnm{J.}\binits{J.}}
(\byear{2006}).
\btitle{Photometric redshifts for the Dark Energy Survey}.
In \bbooktitle{Bulletin of the American Astronomical Society}
\bvolume{38}
\bpages{140}.
\bptok{imsref}%
\end{binproceedings}
\endbibitem

\bibitem[\protect\citeauthoryear{Paulo}{2005}]{Paulo2005}
\begin{barticle}[mr]
\bauthor{\bsnm{Paulo},~\bfnm{Rui}\binits{R.}}
(\byear{2005}).
\btitle{Default priors for {G}aussian processes}.
\bjournal{Ann. Statist.}
\bvolume{33}
\bpages{556--582}.
\bid{doi={10.1214/009053604000001264}, issn={0090-5364}, mr={2163152}}
\bptok{imsref}%
\end{barticle}
\endbibitem

\bibitem[\protect\citeauthoryear{Perlmutter et~al.}{1999}]{Perlmutter1999}
\begin{barticle}[author]
\bauthor{\bsnm{Perlmutter},~\bfnm{S.}\binits{S.}},
  \bauthor{\bsnm{Aldering},~\bfnm{G.}\binits{G.}},
  \bauthor{\bsnm{Goldhaber},~\bfnm{G.}\binits{G.}},
  \bauthor{\bsnm{Knop},~\bfnm{RA}\binits{R.}},
  \bauthor{\bsnm{Nugent},~\bfnm{P.}\binits{P.}},
  \bauthor{\bsnm{Castro},~\bfnm{PG}\binits{P.}},
  \bauthor{\bsnm{Deustua},~\bfnm{S.}\binits{S.}},
  \bauthor{\bsnm{Fabbro},~\bfnm{S.}\binits{S.}},
  \bauthor{\bsnm{Goobar},~\bfnm{A.}\binits{A.}},
  \bauthor{\bsnm{Groom},~\bfnm{DE}\binits{D.}} \betal{et~al.}
(\byear{1999}).
\btitle{Measurements of [Omega] and [Lambda] from 42 high-redshift supernovae}.
\bjournal{The Astrophysical Journal}
\bvolume{517}
\bpages{565--586}.
\bptok{imsref}%
\end{barticle}
\endbibitem

\bibitem[\protect\citeauthoryear{Pissanetzky}{1984}]{Pissanetzky1984}
\begin{bbook}[mr]
\bauthor{\bsnm{Pissanetzky},~\bfnm{Sergio}\binits{S.}}
(\byear{1984}).
\btitle{Sparse Matrix Technology}.
\bpublisher{Academic Press},
  \baddress{London}.
\bid{mr={0751237}}
\bptok{imsref}%
\end{bbook}
\endbibitem

\bibitem[\protect\citeauthoryear{Riess et~al.}{1998}]{Riess1998}
\begin{barticle}[author]
\bauthor{\bsnm{Riess},~\bfnm{A.~G.}\binits{A.~G.}},
  \bauthor{\bsnm{Filippenko},~\bfnm{A.~V.}\binits{A.~V.}},
  \bauthor{\bsnm{Challis},~\bfnm{P.}\binits{P.}},
  \bauthor{\bsnm{Clocchiatti},~\bfnm{A.}\binits{A.}},
  \bauthor{\bsnm{Diercks},~\bfnm{A.}\binits{A.}},
  \bauthor{\bsnm{Garnavich},~\bfnm{P.~M.}\binits{P.~M.}},
  \bauthor{\bsnm{Gilliland},~\bfnm{R.~L.}\binits{R.~L.}},
  \bauthor{\bsnm{Hogan},~\bfnm{C.~J.}\binits{C.~J.}},
  \bauthor{\bsnm{Jha},~\bfnm{S.}\binits{S.}},
  \bauthor{\bsnm{Kirshner},~\bfnm{R.~P.}\binits{R.~P.}} \betal{et~al.}
(\byear{1998}).
\btitle{Observational evidence from supernovae for an accelerating universe
  and a cosmological constant}.
\bjournal{Astronomical Journal}
\bvolume{116}
\bpages{1009--1038}.
\bptok{imsref}%
\end{barticle}
\endbibitem

\bibitem[\protect\citeauthoryear{Roberts and Rosenthal}{2009}]{Roberts2008}
\begin{barticle}[mr]
\bauthor{\bsnm{Roberts},~\bfnm{Gareth~O.}\binits{G.~O.}} \AND
  \bauthor{\bsnm{Rosenthal},~\bfnm{Jeffrey~S.}\binits{J.~S.}}
(\byear{2009}).
\btitle{Examples of adaptive {MCMC}}.
\bjournal{J. Comput. Graph. Statist.}
\bvolume{18}
\bpages{349--367}.
\bid{doi={10.1198/jcgs.2009.06134}, issn={1061-8600}, mr={2749836}}
\bptnote{check year}%
\bptok{imsref}%
\end{barticle}
\endbibitem

\bibitem[\protect\citeauthoryear{Sacks et~al.}{1989}]{Sacks1989}
\begin{barticle}[mr]
\bauthor{\bsnm{Sacks},~\bfnm{Jerome}\binits{J.}},
  \bauthor{\bsnm{Welch},~\bfnm{William~J.}\binits{W.~J.}},
  \bauthor{\bsnm{Mitchell},~\bfnm{Toby~J.}\binits{T.~J.}} \AND
  \bauthor{\bsnm{Wynn},~\bfnm{Henry~P.}\binits{H.~P.}}
(\byear{1989}).
\btitle{Design and analysis of computer experiments}.
\bjournal{Statist. Sci.}
\bvolume{4}
\bpages{409--435}.
\bid{issn={0883-4237}, mr={1041765}}
\bptnote{check related}%
\bptok{imsref}%
\end{barticle}
\endbibitem

\bibitem[\protect\citeauthoryear{Santner, Williams and
  Notz}{2003}]{Santner2003}
\begin{bbook}[mr]
\bauthor{\bsnm{Santner},~\bfnm{Thomas~J.}\binits{T.~J.}},
  \bauthor{\bsnm{Williams},~\bfnm{Brian~J.}\binits{B.~J.}} \AND
  \bauthor{\bsnm{Notz},~\bfnm{William~I.}\binits{W.~I.}}
(\byear{2003}).
\btitle{The Design and Analysis of Computer Experiments}.
\bpublisher{Springer}, \baddress{New York}.
\bid{mr={2160708}}
\bptok{imsref}%
\end{bbook}
\endbibitem

\bibitem[\protect\citeauthoryear{Shaby and Wells}{2011}]{Shaby2011}
\begin{bmisc}[author]
\bauthor{\bsnm{Shaby},~\bfnm{B.}\binits{B.}} \AND
  \bauthor{\bsnm{Wells},~\bfnm{M.~T.}\binits{M.~T.}}
(\byear{2011}).
\bhowpublished{Exploring an adaptive Metropolis algorithm.
Technical Report 2011-14,
Dept. Statistical Science, Duke Univ., Durham, NC.}
\bptok{imsref}%
\end{bmisc}
\endbibitem

\bibitem[\protect\citeauthoryear{Stein}{2008}]{Stein2008}
\begin{barticle}[mr]
\bauthor{\bsnm{Stein},~\bfnm{Michael~L.}\binits{M.~L.}}
(\byear{2008}).
\btitle{A modeling approach for large spatial datasets}.
\bjournal{J. Korean Statist. Soc.}
\bvolume{37}
\bpages{3--10}.
\bid{doi={10.1016/j.jkss.2007.09.001}, issn={1226-3192}, mr={2420389}}
\bptok{imsref}%
\end{barticle}
\endbibitem

\bibitem[\protect\citeauthoryear{Stein, Chi and Welty}{2004}]{Stein2004}
\begin{barticle}[mr]
\bauthor{\bsnm{Stein},~\bfnm{Michael~L.}\binits{M.~L.}},
  \bauthor{\bsnm{Chi},~\bfnm{Zhiyi}\binits{Z.}} \AND
  \bauthor{\bsnm{Welty},~\bfnm{Leah~J.}\binits{L.~J.}}
(\byear{2004}).
\btitle{Approximating likelihoods for large spatial data sets}.
\bjournal{J. R. Stat. Soc. Ser. B Stat. Methodol.}
\bvolume{66}
\bpages{275--296}.
\bid{doi={10.1046/j.1369-7412.2003.05512.x}, issn={1369-7412}, mr={2062376}}
\bptok{imsref}%
\end{barticle}
\endbibitem

\bibitem[\protect\citeauthoryear{Tang}{1993}]{Tang1993}
\begin{barticle}[mr]
\bauthor{\bsnm{Tang},~\bfnm{Boxin}\binits{B.}}
(\byear{1993}).
\btitle{Orthogonal array-based {L}atin hypercubes}.
\bjournal{J. Amer. Statist. Assoc.}
\bvolume{88}
\bpages{1392--1397}.
\bid{issn={0162-1459}, mr={1245375}}
\bptok{imsref}%
\end{barticle}
\endbibitem

\bibitem[\protect\citeauthoryear{Welch et~al.}{1992}]{Welch1992}
\begin{barticle}[author]
\bauthor{\bsnm{Welch},~\bfnm{W.~J.}\binits{W.~J.}},
  \bauthor{\bsnm{Buck},~\bfnm{R.~J.}\binits{R.~J.}},
  \bauthor{\bsnm{Sacks},~\bfnm{J.}\binits{J.}},
  \bauthor{\bsnm{Wynn},~\bfnm{H.~P.}\binits{H.~P.}},
  \bauthor{\bsnm{Mitchell},~\bfnm{T.~J.}\binits{T.~J.}} \AND
  \bauthor{\bsnm{Morris},~\bfnm{M.~D.}\binits{M.~D.}}
(\byear{1992}).
\btitle{Screening, predicting, and computer experiments}.
\bjournal{Technometrics}
\bvolume{34}
\bpages{15--25}.
\bptok{imsref}%
\end{barticle}
\endbibitem

\bibitem[\protect\citeauthoryear{Wikle}{2010}]{Wikle2010}
\begin{bincollection}[mr]
\bauthor{\bsnm{Wikle},~\bfnm{Christopher~K.}\binits{C.~K.}}
(\byear{2010}).
\btitle{Low-rank representations for spatial processes}.
In \bbooktitle{Handbook of Spatial Statistics}
(\beditor{\bfnm{A.~E.}\binits{A.~E.}~\bsnm{Gelfand}},
\beditor{\bfnm{P.}\binits{P.}~\bsnm{Diggle}},
\beditor{\bfnm{M.}\binits{M.}~\bsnm{Fuentes}} \AND
\beditor{\bfnm{P.}\binits{P.}~\bsnm{Guttorp}}, eds.)
\bpages{107--118}.
\bpublisher{CRC Press}, \baddress{Boca Raton, FL}.
\bid{mr={2730946}}
\bptok{imsref}%
\end{bincollection}
\endbibitem

\end{thebibliography}
\end{document}